\documentclass{ws-ijmpa}
\usepackage[super,compress]{cite}
\usepackage{graphicx}
\usepackage[usenames]{color}
\usepackage{colortbl}
\usepackage{epsfig, epstopdf}
\usepackage{hyperref}

\begin{document}
	
\markboth{R.~T.~Ovsiannikov, A.~Yu.~Korchin}{Decay of the Higgs boson to longitudinally polarized leptons and quarks}

%
\catchline{}{}{}{}{}
%

\title{Decay of the Higgs boson to longitudinally polarized leptons and quarks}

\author{R.~T.~Ovsiannikov}

\address{V.N.~Karazin Kharkiv National University, 61022 Kharkiv, Ukraine\\
	NSC 'Kharkov Institute of Physics and Technology', 61108 Kharkiv, Ukraine\\
	roman.ovsiannikov@student.karazin.ua}

\author{A.~Yu.~Korchin}

\address{V.N.~Karazin Kharkiv National University, 61022 Kharkiv, Ukraine\\
	NSC 'Kharkiv Institute of Physics and Technology', 61108 Kharkiv, Ukraine\\
	Institute of Physics, Jagiellonian University, Lojasiewicza 11, 30-348 Krakow, Poland\\
korchin@kipt.kharkov.ua}

\maketitle

\begin{history}
	\received{16 January 2024}
	\revised{27 March 2024}
	\accepted{31 March 2024}
\end{history}

 
\begin{abstract}
An important problem in the particle physics is interaction of the Higgs boson with the fermions. It is in the processes, which involve $h f f $ interaction, the manifestations of $CP$ violation are possible. This in turn can be helpful in solving the problem of the dominance of matter over antimatter in the Universe. In this connection, in the present paper, the effect of longitudinal polarization of leptons and quarks produced in the decay of the Higgs boson is investigated and calculated.  We consider the case in which the Higgs boson interacts with fermions via a mixture of scalar and pseudoscalar couplings. Under this assumption, the longitudinal polarization can acquire non-zero values due to imaginary part of loop corrections to tree-level amplitudes. This polarization is a direct signature of the $CP$ violation in the Higgs sector.  Effects of the longitudinal polarization of $\tau$ leptons  
on the energy distribution of the pions in the two-step Higgs-boson decay $h \to \tau^- \tau^+ \to 
\pi^- \nu_\tau \pi^+ \bar{\nu}_\tau$ are also studied, and the energy asymmetry with corresponding moments are calculated.

\keywords{Higgs boson decay, polarized fermions, one-loop vertex corrections}   
\end{abstract}

\ccode{PACS numbers:11.30.Er, 12.15.Ji}

\maketitle
\setcounter{footnote}{0}

\section{Introduction}

The important problem of Higgs boson \cite{lib1, lib2}, discovered in 2012, is its interaction with 
particles of matter - leptons and quarks. In the Standard Model (SM) the Higgs boson $h$ has positive $CP$ parity, i.e. it is a $CP$-even scalar particle. At the same time, many models beyond the SM contain scalar particles that have positive or negative $CP$-parity. There may also be particles that have no definite parity 
(see e.g. \cite{lib3}), being the mixture of the $CP$-even and $CP$-odd states.  
The elucidation of such properties of the discovered boson $h$ is also an essential element in the study of the 
mechanism of the electroweak symmetry breaking and the generation of particle masses.

A problem related to $CP$-symmetry is the matter-antimatter asymmetry in the Universe. The $CP$-violation, which occurs in the SM due to the Cabibbo-Kobayashi-Maskawa matrix (CKM) \cite{lib4, lib5}, describes all flavor changing  physical processes. However, theoretical calculations \cite{lib6, lib7} show that this violation is too small to explain the observed asymmetry of matter and antimatter in the Universe. Therefore, there should possibly be other sources of the $CP$ violation beyond the CKM mechanism. Thus, the search for additional sources of $CP$-violation is one of the important directions in the high-energy physics. 
And one possibility in this quest is to investigate the decay of the Higgs boson into polarized leptons and quarks.

In the recent years, some of decays $h \to f \bar{f}$ have been measured experimentally at the Large Hadron 
Collider (LHC). But more information can be expected from experiments at the planned future electron-positron colliders, such as ILC and CLIC, which will aim at studying in detail the properties of the Higgs boson and the heaviest elementary particle -- top quark. The present status of the LHC measurements of the Higgs boson couplings to  leptons, quarks and vector bosons, and their CP properties, is reviewed in Ref.~\cite{Bass:2021}

It should be noted that measurements at the LHC of the Higgs-boson decay width  (or decay rate) 
$\Gamma (h \to f \bar{f})$ are not sensitive to possible $CP$-violation effects. The LHC experiments at the CMS and ATLAS~\cite{CMS:2012vby, CMS:2014nkk, ATLAS:2013xga} confirm the spin and parity of the 
SM Higgs boson.  
In addition, constraints on the mass of pseudoscalar Higgs boson in the two-Higgs-doublet models have been obtained by the CMS collaboration~\cite{CMS:2019rvj}. ATLAS experiment~\cite{ATLAS:2020ior} constrained the $CP$ mixing angle $\phi_{PS}$ in the Yukawa $h t t$ coupling and excluded the interval  
$|\phi_{PS}|> 43^\circ$ at 95\% confidence level. In more recent CMS experiments~\cite{CMS:2021nnc} the following constraints on the $CP$-even and $CP$-odd   
$h t t$ couplings, $a_t = 1.05^{+0.25}_{-0.20}$ and $b_t = -0.01^{+0.69}_{-0.67}$, were obtained. 
These results indicate that the coupling $b_t$ is close to zero with large uncertainties. Based on these results  
we here assume that in the $h f f $ interaction the coupling 
$b_f$ is very small, and therefore the width of the decay $h  \to f \bar{f}$  
for unpolarized fermions is not sensitive to $b_f$, because it is proportional to $|b_f |^2$ 
(see Eq.~(\ref{eq2}) below). 
Thus it would be interesting to study observables which are linear in the $CP$-violating parameter $b_f$. 

The present paper deals with the decay of the scalar boson $h$ into polarized leptons and quarks. 
The idea is to enhance the sensitivity to the small effect of the $CP$-odd state by measuring the longitudinal polarization of the final fermions.  
The suitable decay to study the fermion polarizations is, for example, the Higgs decay to the $\tau$ leptons 
because of the hadronic decay channels of the $\tau$ lepton. We should mention, that decay of 
the Higgs boson into $\tau$-leptons with their subsequent hadronic decays was studied in 
Ref.~\cite{Berge:2011ij}, although the consideration there was carried out on the tree level 
in which polarization of $\tau$ lepton is zero. Various aspects of the $h \to \tau^- \tau^+ $ decay 
with emphasis on $CP$ violation effects and application of machine learning techniques were studied  
in Refs.~\cite{Bower:2002zx, Desch:2003mw, Desch:2003rw, Barberio:2017ngd, Lasocha:2020ctd}.

In our paper, the first-order loop corrections to $h \to \tau^- \tau^+$   
are calculated; they give rise to a non-zero longitudinal polarization of $\tau$ leptons which is linear in the small 
$CP$-odd parameter.
This polarization is reflected in the energy distribution of the $\pi$ mesons, produced in the 
decays of the $\tau$ leptons. The distribution of the pions can in principle be studied experimentally.     

The following sections introduce the necessary formalism to calculate the polarization of the final quarks and leptons. Note that some aspects of this problem have been investigated earlier in~\cite{Bernreuther:1997, lib8}.

\section{Higgs boson interaction with fermions}
\label{sec:interaction vertex}

Consider the decay of the Higgs boson into a fermion-antifermion pair, $h \rightarrow f \bar f$.  
In particular, consider these decays in the Higgs-boson rest frame, and use $\hbar = c = 1$.

In the SM the $h f \bar{f}$ interaction Lagrangian is a scalar and a Hermitian operator. 
Assuming that there is also a pseudoscalar (and under some assumptions a non-Hermitian) part of this 
interaction, the Lagrangian reads as follows:
\begin{equation}\label{eq1} 
\mathcal{L} = -\dfrac{m_f}{v} h \, \bar \psi_f (a_f + i b_f \gamma_5) \psi_f,
\end{equation}
where $a_f$ and $b_f$ are complex parameters, such as $|a_f|^2 + |b_f|^2 = 1$ ($a_f = 1$ and $b_f = 0$ correspond to the SM), $m_f$ is the mass of the fermion $f$, $v = \left (\sqrt{2}G_f \right)^{-1/2} \approx 246$ $GeV$,  and $G_F \approx 1.1663 \cdot 10^{-5}$ $GeV^{-2}$ is the Fermi constant \cite{PDG}. 

First, the width of the decay to non-polarized fermions is 
\begin{equation}\label{eq2} 
\Gamma(h \rightarrow f \bar f) = \dfrac{N_f G_f}{4 \sqrt{2} \pi} m_f^2 m_h \beta_f 
\left(|a_f|^2 \beta_f^2 + |b_f|^2 \right), 
\end{equation}
where $N_f = 1$ for leptons and $N_f = 3$ for quarks, $\beta_f = \sqrt{1 - {4 m_f^2}/{m_h^2}}$ 
is the velocity of fermions in the Higgs-boson rest frame, and $m_h \approx 125.26$ GeV \cite{PDG}. 
In general, if a pseudoscalar part of the interaction exists, its contribution is substantially smaller than that of the scalar part. Indeed, experiments at the LHC put strong constraints on value of the pseudoscalar 
coupling~\cite{CMS:2021nnc}. However the decay width (\ref{eq2}) is not convenient for measuring the pseudoscalar admixture since the width is quadratic in the parameter $b_f$ (note that $\beta_f \approx 1$). The situation would be more favorable if some observables were linear in the coupling  $b_f$. 
\begin{figure}[h]  

\centerline{\includegraphics[width=8cm]{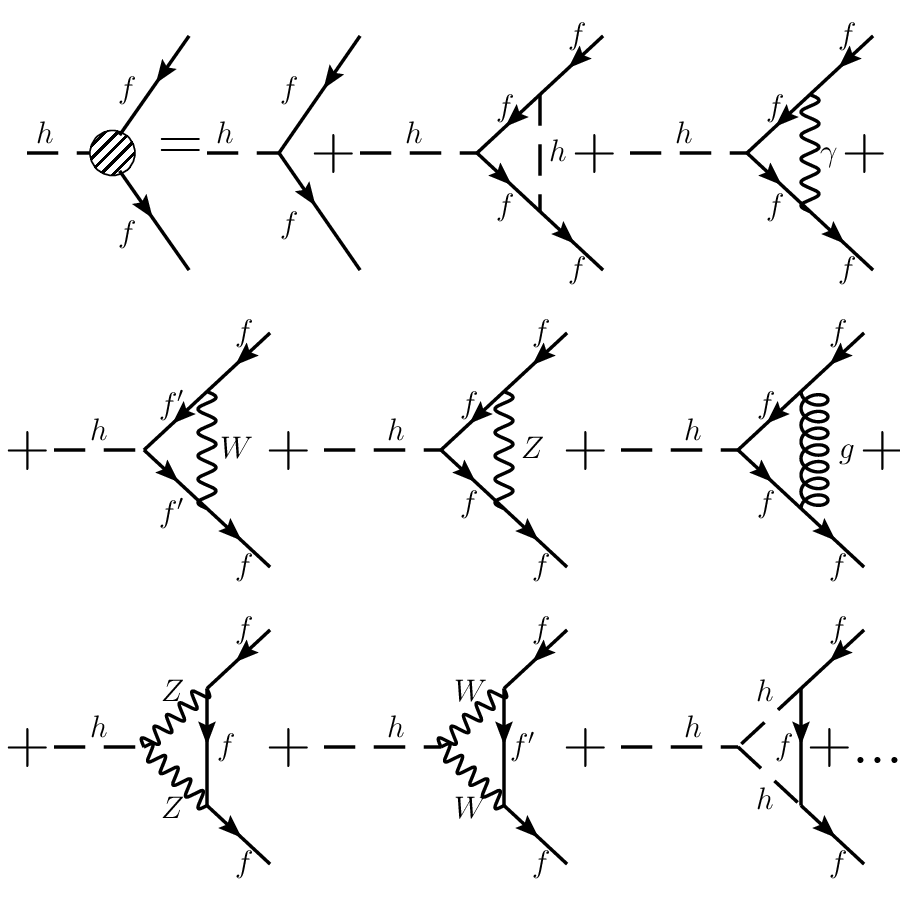}}
\caption{Modification of the Higgs-boson interaction with a fermion 
$f = (\ell, \, q)$ due to one-loop vertex corrections. 
In case of decay into quarks, $f^\prime$ is a quark with the electric charge 
$e Q_{f^\prime}=(1/3 - Q_f) e$, if the charge of the quark $f$ is $e Q_f$. 
In the case of decay into charged leptons, $f^\prime$ is a neutrino.  
Note also that the gluon exchange (the 6th term of the sum) is present only in the decay to quarks.}
\label{ris:Figure_1}
\end{figure}

Let us now consider the decay of the Higgs boson to the polarized fermion and antifermion which are described  by the polarization vectors in their corresponding rest frames, $\vec{\xi}_1$ and $\vec{\xi}_2$.
The decay width now takes the form:
\begin{eqnarray} \label{eq3} 
\dfrac{\partial \Gamma}{\partial\Omega} &=& \dfrac{\Gamma(h \rightarrow f \bar f)}{16 \pi}
 \Bigl\{1 - \xi_{1L} \xi_{2L}  
- \dfrac{2 \, {\rm Re}(a_f b_f ^*)}{|a_f|^2 \beta_f^2 + |b_f|^2}\beta_f \vec{n} \cdot 
[\vec{\xi}_{1T} \times \vec{\xi}_{2T}]  \nonumber \\ 
&&+ \dfrac{|a_f|^2 \beta_f^2 - |b_f|^2}{|a_f|^2 \beta_f^2 + |b_f|^2} 
(\vec{\xi}_{1T} \cdot \vec{\xi}_{2T})  
- \dfrac{2 \, {\rm Im}(a_f b_f^*)}{|a_f|^2 \beta_f^2 + |b_f|^2} \beta_f (\xi_{1L} - \xi_{2L}) \Bigr\},
\end{eqnarray}
where  
$\vec{n}$ is a unit vector in the direction of the 3-momentum of the fermion in the Higgs-boson rest frame; 
$\xi_{1L,2L} \equiv (\vec{n} \cdot \vec{\xi}_{1,2})$ are the longitudinal polarization components, $\vec{\xi}_{1T,2T} 
\equiv \vec{\xi}_{1,2}-\vec{n}(\vec{n} \cdot \vec{\xi}_{1,2})$ are the transverse polarization components. 
As it is seen from (\ref{eq3}), the linear terms $\propto b_f$ indeed appear, and moreover, there is
a non-zero longitudinal polarization of the antifermion/fermion
\begin{equation} 
\label{eq4}
\mathcal{P}^{(0)} =  \pm \dfrac{{2 \, \rm Im}(a_f b_f^*)}{|a_f|^2 \beta_f^2 + |b_f|^2}\beta_f.
\end{equation}
For definiteness, we discuss below the polarization of the antifermions.

Of course, in the SM the interaction is Hermitian and $a_f$ and $b_f$ are real parameters. Hence the longitudinal polarization at the tree level is zero. However an imaginary part of scalar and pseudoscalar parameters is 
generated because of the loop vertex corrections, or radiative corrections. These radiative corrections do not change the structure of the vertex in Eq.~(\ref{eq1}), so that the tree level and the radiative corrections together 
form an effective $h f \bar{f}$ vertex (Fig.~\ref{ris:Figure_1})
\begin{equation} \label{eq5}
\widetilde{V}_{h f \bar f} = -i \dfrac{m_f}{v} \left(A_f + i B_f \gamma_5 \right).
\end{equation}

Let us denote $A_f = a_f + \Delta A_f$, $B_f = b_f + \Delta B_f$, where $\Delta A_f$ and $\Delta B_f$ are complex numbers representing the radiative corrections. We assume  that these corrections are small, so that     
$|\Delta A_f| \ll a_f$ and $|\Delta B_f| \ll b_f$. 
Thus one immediately obtains the longitudinal polarization in the first order in $\Delta A_f$ and $\Delta B_f$:
\begin{equation}\label{eq6}
\mathcal{P}^{(1)} =   2\beta_f \dfrac{-a_f \, 
{\rm Im} (\Delta B_f) + b_f \, {\rm Im} (\Delta A_f)}{|a_f|^2 \beta_f^2 + |b_f|^2}.
\end{equation}

For further convenience let us introduce the ``pseudoscalar angle'' $\phi_{PS}$ defined as  
\begin{equation} 
\tan{\phi_{PS}} \equiv \dfrac{b_f}{a_f}.
\end{equation}

\section{Radiative corrections}
\label{sec:radiative corrections }

We should note that radiative corrections to the decay width $\Gamma(h \to f \bar{f})$ have been 
calculated in Refs.~\cite{Braaten:1980, Janot:1989, Drees:1990, Kalyniak:1991}. 
Advanced studies of the higher-order QCD corrections to the decay width of the Higgs boson to quarks and hadrons have been 
performed in Refs.~\cite{Gorishnii:1990zu, Gorishnii:1991zr, Kataev:1993be, Kataev:2008ntk}.
Such calculations require accurate treatment of ultraviolet and infrared 
divergences as well as renormalization.  
However, in the consideration of the longitudinal polarization, as it is seen from  Eq.~(\ref{eq6}), 
we need only the imaginary part of the matrix elements. This considerably simplifies calculations.    
One can use Cutkosky's  rules 
(see, {\it e.g.} \cite{lib9}) which imply that intermediate fermions in the loops are on-mass-shell.   
This procedure reduces the integration area in the four-dimensional integrals and makes them finite.  

\begin{figure}[h]  
	\centerline{\includegraphics[width=10cm]{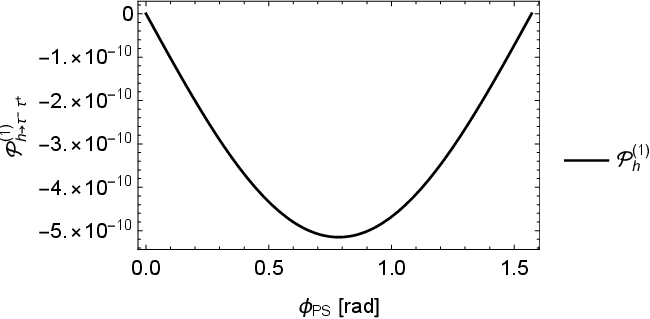}}
	\caption{Longitudinal polarization of final $\tau$-leptons, appearing from an intermediate 
	Higgs boson in Fig.~\ref{ris:Figure_1}, vs. the pseudoscalar angle $\phi_{PS}$.}
	\label{fig_2}
\end{figure}

\begin{figure}[h]  
	\centerline{\includegraphics[width=10cm]{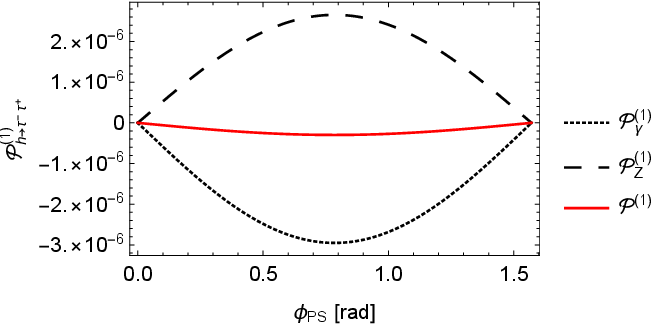}}
	\caption{Longitudinal polarization of final $\tau$-leptons, appearing from corrections from the photon and $Z$-boson exchanges in Fig.~\ref{ris:Figure_1}, vs.  the angle $\phi_{PS}$.
The red line shows dependence of the total longitudinal polarization on $\phi_{PS}$.}
	\label{fig_3}
\end{figure}

There are two points to mention. The last three diagrams in Fig.~\ref{ris:Figure_1} do not contribute to the imaginary parts of $A_f$ and $B_f$ since the intermediate $Z, \, W^\pm$ and $h$ bosons cannot be on their mass shells.   
Another point is that the corrections with the photon and gluon exchanges are infrared divergent due to the 
zero mass of the gluon and the photon. This divergence is preserved in the imaginary part of the matrix element. 
In general, this divergence is canceled if one accounts for the processes 
$h \rightarrow \tau^+ \tau^- \gamma$ with real soft photon, and $h \rightarrow q \bar q g$ with real soft gluon.  
However, in calculation of the longitudinal polarization it turns out that the 
infrared divergencies in the combination $-a_f \, 
{\rm Im} (\Delta B_f) + b_f \, {\rm Im} (\Delta A_f)$ in the numerator of Eq.~(\ref{eq6}) 
cancel, and the value remains finite.

Now let us calculate each contribution to the longitudinal polarization of the final fermions. 
In the decay to leptons, there are three such contributions: 
2nd, 3rd, and 5th diagrams in the sum in Fig. \ref{ris:Figure_1}. 
We do not include the fourth term of the sum (Fig.~\ref{ris:Figure_1}) because 
it involves the Higgs-neutrino interaction vertex.
While in the decay to quarks, we have five contributions: the 2nd, 3rd, 4th, 
5th, and 6th diagrams in the sum in Fig. \ref{ris:Figure_1}. 
  
Then we find for the 2nd diagram in Fig. \ref{ris:Figure_1}:
\begin{equation}\label{eq8}
	\mathcal{P}^{(1)}_h = - \dfrac{m_f^2 \sin{2 \phi_{PS}}}{2 \pi m_h^2 (\beta_f^2 \cos^2{\phi_{PS}}  + \sin^2{\phi_{PS}})} \left(\dfrac{m_f}{v}\right)^2 
	\Bigl[1 - \dfrac{1}{\beta_f^2} \log \left(1 + \beta_f^2 \right) \Bigr].
\end{equation}

For the 3rd diagram in Fig. \ref{ris:Figure_1} we obtain:
\begin{equation}\label{eq9}
	\mathcal{P}^{(1)}_\gamma = - \dfrac{m_f^2 Q_f^2 e^2 \sin{2 \phi_{PS}}}{2 \pi m_h^2 (\beta_f^2 \cos^2{\phi_{PS}}  + \sin^2{\phi_{PS}})}.
\end{equation}
Here $Q_f$ is the charge of the final fermion in units of the positron charge $e$. 
The 4th diagram in the sum in Fig. \ref{ris:Figure_1} contributes to the polarization only in 
the $h$ decays to quarks. This contribution has the following form:
\begin{multline}
\mathcal{P}^{(1)}_W =  -\dfrac{ g^2_W \sin{2 \phi_{PS}}}{16 \pi \beta_f^2 m_h^4 m_W^2(\beta_f^2 \cos^2{\phi_{PS}}  + \sin^2{\phi_{PS}})}  
\\ \times  \sum_{i} | V_{f i}|^2  m_i^2 
\Biggl\{ (k_+^2 -k_-^2)\Big[2 m^2_W - m_h^2\beta_f^2 + m_i^2 - m_f^2\Big] -\\- \Big[(2 m^2_W + m_i^2 - m_f^2)(m_f^2 - m_i^2 + m^2_W)  
+ m_h^2 m_W^2 \beta_f^2\Big] \log \Bigl(\dfrac{k_+^2+m_W^2}{k_-^2+m_W^2}\Bigr) \Biggr\}.
\end{multline}
Here $g_W={e}/{s_w}$, $s_w=\sin \Theta_w$,  $\Theta_w=\arccos ({m_W}/{m_Z})$ is weak mixing angle 
(Weinberg angle); $m_W \approx 80.39$ GeV, $m_Z \approx 91.19$ GeV \cite{PDG}; the sum runs over intermediate quarks $i$ with the masses $m_i \leq m_h/2$,  $V_{i f}$ is the element of the CKM matrix, and
\begin{equation}
k_{\pm} = 
	\begin{cases}
	\pm \dfrac{m_h \beta_f}{2} + \sqrt{\dfrac{m_h^2 \beta_f^2}{4} + m_f^2 - m_i^2}, \qquad m_f > m_i \\
	\dfrac{m_h \beta_f}{2} \pm \sqrt{\dfrac{m_h^2 \beta_f^2}{4} + m_f^2 - m_i^2}, \qquad m_f < m_i.
	\end{cases}
\end{equation}

For the 5th diagram in (Fig. \ref{ris:Figure_1}) the result is:
\begin{eqnarray}\label{eq12}
	\mathcal{P}^{(1)}_Z &=&  \dfrac{m_f^2 g_Z^2 \sin{2 \phi_{PS}}}{8 \pi m_h^2 (\beta_f^2 \cos^2{\phi_{PS}}  + \sin^2{\phi_{PS}})} \Biggl\{ a_f^2 \Bigl[ \beta_f^2 \dfrac{m_h^2}{m_Z^2} + \log \Bigl(1 + \beta_f^2 \dfrac{m_h^2}{m_Z^2} \Bigr) \Bigr] \nonumber \\ 
	&-& (a^2_f + v^2_f) \Bigl[1 -  \dfrac{m_Z^2}{m_h^2 \beta_f^2} 
	\log \Bigl(1 + \beta_f^2 \dfrac{m_h^2}{m_Z^2} \Bigr) \Bigr]  \Biggr\},
\end{eqnarray}
where $g_Z={e}/({s_w c_w})$, $c_w=\cos \Theta_w$,  $v_f$ and $a_f$ are respectively vector and axial-vector coupling constants for a fermion $f$. 

The longitudinal polarization of the quarks due to the gluon exchange (the 6th diagram in Fig. \ref{ris:Figure_1}) is:
\begin{equation}
	\mathcal{P}^{(1)}_g = - \dfrac{2 m_q^2 g^2 \sin{2 \phi_{PS}}}{3 \pi m_h^2 (\beta_q^2 \cos^2{\phi_{PS}}  + \sin^2{\phi_{PS}})},
\end{equation}
where $m_q$ is the mass of the final quark, 
$\beta_q = \sqrt{1 - {4 m_q^2}/{m_h^2}}$ is the velocity of quarks in the Higgs-boson rest frame. 
Also here 
\begin{equation}
\dfrac{g^2}{4 \pi} = \dfrac{12 \pi}{(33 - 2 n_f) \log \left({m_h^2}/{\Lambda^2} \right)}
\end{equation}
is the running strong coupling constant \cite{lib10},  $n_f$ is an effective number of quark flavors 
with masses less than $m_h/2$, and $\Lambda \approx 200-300$ GeV is the QCD scale parameter. 

It can be seen that the longitudinal polarization in each case strongly depends on the mass of the final fermions, approximately quadratically. This means that to increase the longitudinal polarization one has to consider processes involving the heaviest leptons and quarks. The $\tau$-lepton ($m_\tau \approx 1.78$ GeV) and $b$-quark ($m_b \approx 4.67$ GeV) are thus the most suitable fermions. 

\begin{figure}[h!]  
	\centerline{\includegraphics[width=10cm]{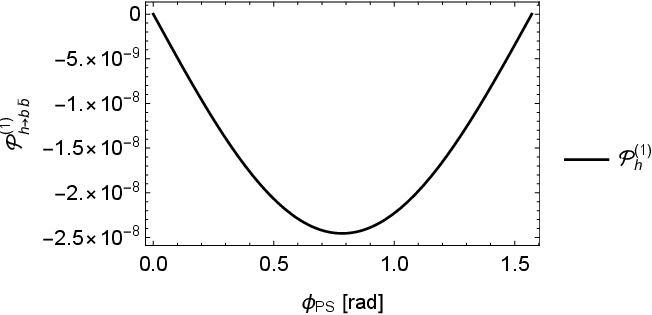}}
	\caption{Longitudinal polarization of $b$-quarks appearing from a correction for the Higgs interaction.}
	\label{fig_4}
\end{figure}

\begin{figure}[h!]  
	\centerline{\includegraphics[width=10cm]{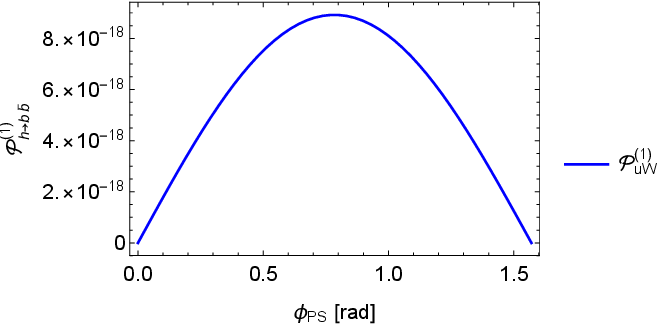}}
	\caption{Longitudinal polarization of $b$-quarks appearing from a correction for 
	the weak interaction (exchange of the $W$-boson with the intermediate $u$-quark).}
	\label{fig_5}
\end{figure}

\begin{figure}[h!]  
	\centerline{\includegraphics[width=10cm]{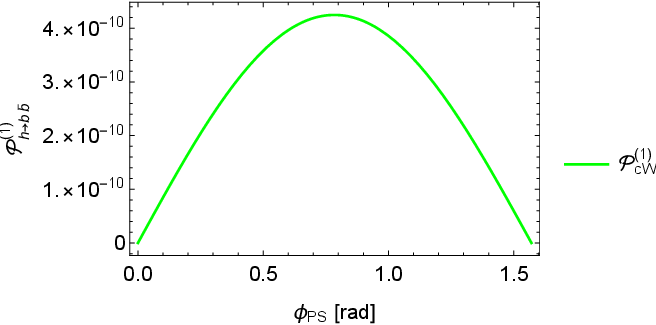}}
	\caption{Longitudinal polarization of $b$-quarks appearing from a correction for the weak interaction (exchange of the $W$-boson with the intermediate $c$-quark).}
	\label{fig_6}
\end{figure}

\begin{figure}[h]  
	\centerline{\includegraphics[width=10cm]{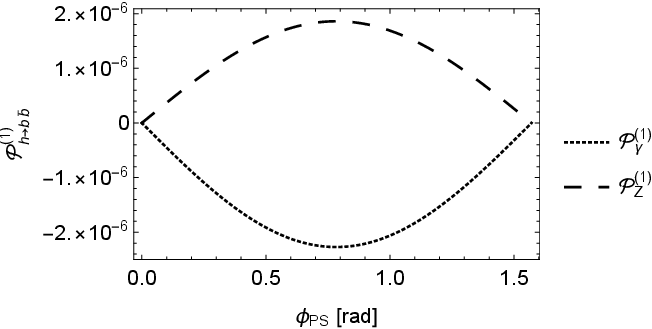}}
	\caption{Longitudinal polarization of final $b$-quarks appearing from electromagnetic and weak (exchange of the $Z$-boson) interaction corrections.}
	\label{fig_7}
\end{figure}

\begin{figure}[h!]  
	\centerline{\includegraphics[width=10cm]{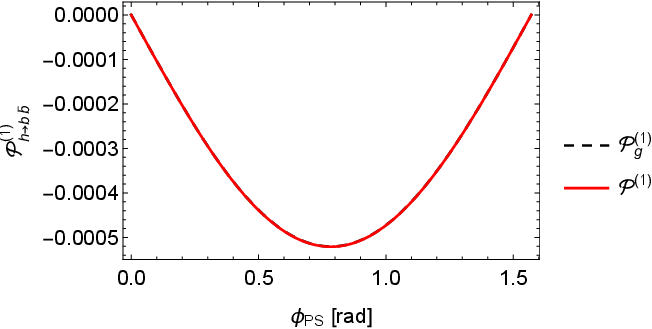}}
	\caption{Longitudinal polarization of the $b$-quarks appearing from the correction for the strong interaction. The red line indicates the dependence of the total longitudinal polarization on the pseudoscalar parameter of the $h f \bar{f}$ interaction. It is seen that the main contribution comes from the gluon interaction.}
	\label{fig_8}
\end{figure}

Let us discuss the results presented in Figs.~\ref{fig_2}--\ref{fig_8}. As can be seen, the main contribution to the longitudinal polarization, in the case of $\tau$-leptons, comes from corrections due to due to the weak and electromagnetic interactions (Fig. \ref{fig_3}), which cancel to the large extend, because these corrections have opposite signs.    

The similar situation can be traced to the polarization of the $b$-quark (Fig. \ref{fig_7}). However, it turns out that here the main contribution is due to a correction for the strong interaction (Fig. \ref{fig_8}). 
Also, in both cases, the contribution from the virtual Higgs boson is significantly smaller 
(Fig. \ref{fig_2} and Fig. \ref{fig_4}), compared to the above corrections. 
In the case of $b$-quarks, there are also two very small corrections due to the $W$-boson exchange 
(Figs. \ref{fig_5} and \ref{fig_6}). Due to the small mass of the intermediate $u$-quark, the contribution from this diagram is negligible (Fig. \ref{fig_5}). However, with the intermediate $c$-quark  the contribution from this 
diagram (Fig. \ref{fig_6}) is comparable with that from the Higgs-boson 
exchange (Fig. \ref{fig_4}), although it is still very small. 
We conclude that for observation of the longitudinal polarization of fermions in the Higgs boson decay
the most suitable final state is the $b \bar{b}$ pair.

 
In experiments, the polarization of the $\tau$-lepton is difficult to measure directly 
because of a very short lifetime $\tau = (290.3 \pm 0.5) \times 10^{-15}$ s \cite{PDG}. An indirect method of determining the polarization is to measure the azimuthal angle distribution of the $\pi$-mesons in the $h \rightarrow \tau^- \tau^+ \rightarrow \pi^- \nu_\tau \pi^+ \bar \nu_\tau$ decay \cite{lib11}. 

\section{Decay of the Higgs boson $h \rightarrow \tau^- \tau^+ \rightarrow \pi^- \nu_\tau \pi^+ \bar \nu_\tau$}
\label{sec:pions}

Let us take a closer look at the decay channel 
$h \rightarrow \tau^- \tau^+ \rightarrow \pi^- \nu_\tau \pi^+ \bar \nu_\tau$. Using the distribution (\ref{eq3}) and argumentation of Refs.~\cite{Kawasaki:1973, Grzadkowski:1995}, we can obtain expression for the distribution of the decay width over the energies of the secondary $\pi^\pm$-mesons: 
\begin{equation}
\dfrac{ d^2 \Gamma(h \rightarrow \tau^- \tau^+ \rightarrow \pi^- \nu_\tau \pi^+ \bar \nu_\tau)}{ d \omega \,
d  \omega'} 
= \Gamma(h \rightarrow \tau^- \tau^+) 
  \Bigl({\rm BR}(\tau \rightarrow \pi \nu_\tau) \Bigr)^2 \dfrac{ d^2 W}{d  \omega \, d  \omega'}, 
\end{equation}
where ${\rm BR}(\tau \rightarrow \pi \nu_\tau)$ is the branching of the $\tau$ decay 
through the $\tau \rightarrow \pi \nu_\tau$ channel, $\dfrac{d^2 W}{d \omega \, d \omega'}$ 
is the distribution of the Higgs-boson decay, normalized to unity, as a function of $\pi^-$  
and $\pi^+$ energies, $\omega$ and $\omega^\prime$, respectively, in the Higgs-boson rest frame. 
This distribution has the form:
\begin{equation}
\dfrac{d^2 W}{d \omega \, d \omega'} = \dfrac{1}{4 \big(\tilde{\omega} \beta_\tau \beta_{\pi} \gamma_\tau\big)^2} \Biggl\{1 - \mathcal{P}_\tau \dfrac{\big(\omega - \omega'\big)}{\tilde{\omega} \beta_{\pi} \gamma_\tau \beta_\tau} -   
\dfrac{1}{\big(\beta_\pi \beta_\tau \gamma_\tau\big)^2} \Big(\dfrac{\omega}{\tilde{\omega}} - \gamma_\tau \Big) \Big(\dfrac{\omega'}{\tilde{\omega}} - \gamma_\tau \Big) \Biggr\}.
\label{eq:energy}
\end{equation} 
where $\gamma_\tau = \dfrac{m_h}{2 m_\tau}$ (in the Higgs boson rest frame), $\beta_\pi = \dfrac{m_\tau^2 - m_\pi^2}{m_\tau^2 + m_\pi^2}$ is the velocity of the $\pi$-mesons in the rest frame of $\tau^\mp$ lepton, 
$\tilde{\omega} = \dfrac{m_\tau^2 + m_\pi^2}{2 m_\tau}$ is the energy of the $\pi^\mp$ mesons in 
the $\tau^\mp$ rest frame, and $m_\pi = 139.6$ MeV; 
$\mathcal{P}_\tau = \mathcal{P}_h^{(1)} + \mathcal{P}_Z^{(1)} + \mathcal{P}_\gamma^{(1)}$, expressions for polarizations are taken from (\ref{eq8}), (\ref{eq9}) and (\ref{eq12})  (with the index $f$ replaced by $\tau$). 
The energy of $\pi$-mesons varies between the minimal and maximal values
\begin{equation}
\omega_{min/max} = \tilde{\omega} \gamma_\tau (1 \mp \beta_\tau \beta_\pi). 
\label{eq:omega_min_max}
\end{equation} 
It is seen from (\ref{eq:energy}) that there is energy asymmetry in the decay width under the 
substitution $\omega \leftrightarrow \omega^\prime$. Let us analyze this asymmetry and define it as follows:
\begin{eqnarray} \label{eq17}
A(\omega, \omega') &=& \Big(\dfrac{d^2 W}{d \omega \, d \omega'} - \dfrac{d^2 W}{d \omega' \, d \omega}\Big)\Big/\Big(\dfrac{d^2 W}{d \omega \, d \omega'} + \dfrac{d^2 W}{d \omega' \, d \omega}\Big) \nonumber  \\ 
&=& - \dfrac{\mathcal{P}_\tau}{\tilde{\omega}  \beta_\pi \gamma_\tau \beta_\tau} \cdot 
 \dfrac{\omega - \omega'}{1 - \dfrac{1}{\big(\beta_\pi \beta_\tau \gamma_\tau\big)^2} \Big(\dfrac{\omega}{\tilde{\omega}} - \gamma_\tau \Big) \Big(\dfrac{\omega'}{\tilde{\omega}} - \gamma_\tau \Big)}.
\end{eqnarray}

It is convenient to introduce moments of asymmetry, and normalize them to the mean energy of the $\pi$-meson in order to have the moments dimensionless, as well as the asymmetry:
\begin{eqnarray} \label{eq18} 
A^{(n)} &=& \dfrac{1}{(\tilde{\omega} \gamma_\tau)^{n+2}} \int\limits_{\omega_{min}}^{\omega_{max}} d\omega \int\limits_{\omega_{min}}^{\omega_{max}} d\omega' \, A(\omega, \omega')(\omega - \omega')^n  \nonumber  \\ 
&=& -(\beta_\pi \beta_\tau)^{n+2} \mathcal{P}_\tau \int\limits_{-1}^1 dx \int\limits_{-1}^1 dy\, \dfrac{(x + y)^{n + 1}}{1 + xy} \nonumber \\ 
&\equiv& -(\beta_\pi \beta_\tau)^{n+2} \mathcal{P}_\tau a^{(n)}. 
\end{eqnarray}
The two-dimensional integrals in (\ref{eq18}) can be calculated analytically (see Appendix A). 

\begin{figure}[h]  
	\centerline{\includegraphics[width=10cm]{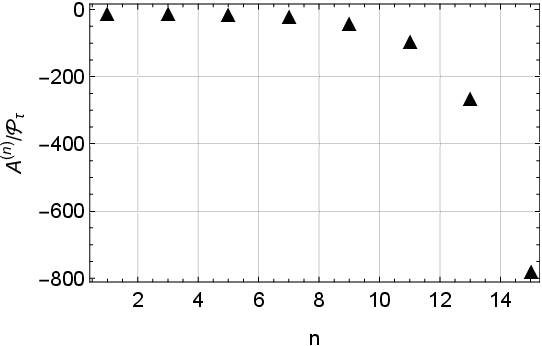}}
	\caption{The first eight odd moments of the energy asymmetry of charged pions in the Higgs boson decay. 
	All values of the moments are divided by the polarization value.}
	\label{fig_9}
\end{figure}

Fig. \ref{fig_9} demonstrates several moments of the energy asymmetry in Eq.~(\ref{eq18}). Apparently all even moments vanish. This asymmetry can be studied experimentally by measuring the numbers 
of the produced $\pi^+$ and $\pi^-$ mesons with the same energies. 
From (\ref{eq17}) and (\ref{eq18}) we see that the asymmetry and its moments are directly proportional to the longitudinal polarization of the $\tau$-leptons. Thus, if it is possible to measure the asymmetry 
of the energy distribution, and its moments (see Fig. \ref{fig_9}), then it would be possible to determine the $\tau$ longitudinal polarization and in this way, the admixture of the pseudoscalar Higgs-fermion interaction. Of course, for the purely scalar $h f \bar{f}$ interaction there will be no polarization, nor will there be energy asymmetry.

\section{Conclusions}
\label{sec:conclusions}

This paper deals with the decay of the Higgs boson into a pair of polarized fermions. 
The longitudinal polarization of the fermion (antifermion) is calculated 
under assumption that the Higgs boson interaction with fermions is a mixture of scalar and pseudoscalar 
couplings. This assumption is based on various models beyond the SM, where the scalar particles can have 
positive or negative $CP$ parity, or no definite $CP$ parity.   

The calculated value of longitudinal polarization appeared to be quite small for leptons, 
while it is several orders of magnitude larger 
for heavy quarks, mainly because of the gluon exchange mechanism. The effect of longitudinal polarization may be significantly enhanced in the decays of other heavy bosons, which are envisaged in the models beyond the SM. 
Any observation of longitudinal polarization in experiments would indicate 
additional source of $CP$ violation in the Higgs-boson sector beyond the CKM mechanism, 
that may be helpful in solution of the problem of matter over antimatter dominance.  

The two-step decay $h \rightarrow \tau^- \tau^+ \rightarrow \pi^- \nu_\tau \pi^+ \bar \nu_\tau$ is also 
considered. It is shown that the pseudoscalar part of the Higgs-fermion interaction gives rise to asymmetry in the distribution of the decay width over the energies of the $\pi^\pm$-mesons.  This observable can be convenient for experimental studies.  

%
\section*{Acknowledgments}

The authors acknowledge partial support by the National Academy of Sciences of Ukraine 
via the program  ``Participation in the international projects in high-energy and 
nuclear physics'' (project C-4/53-2023).   

This project was also supported in part from funds of Polish National Science Centre
under decision  No. UMO-2022/01/3/ST2/00027. 
A.Yu.K. is thankful to the Polish Academy of Sciences for financial support.  

\vspace*{\fill}

\appendix
\section{Moments of asymmetry}
\label{sec:appex}
The analytical expression for the moments $a^{(n)}$ of the energy asymmetry is:
\[
a^{(n)} = 
\begin{cases}
0 \quad if \quad \text{$n$ is even};\\
8\Large \sum\limits_{k=0}^{(n - 3)/4}\dfrac{C_{n+1}^{2k}}{n - 4k + 1} \Large \sum\limits_{l=0}^{(n - 4k - 1)/2} \dfrac{1}{2k + 1 + 2l} - \\ - 8\Large \sum\limits_{k=0}^{(n - 3)/4}\dfrac{C_{n+1}^{2k+1}}{n - 4k - 1} \Large \sum\limits_{l=0}^{(n - 4k - 3)/2} \dfrac{1}{2k + 3 + 2l} - \\ - 4 C_{n+1}^{(n+1)/2} \Large \sum\limits_{l=0}^{(n - 3)/4} \dfrac{1}{(2l+1)^2} + \\ + \dfrac{\pi^2}{2} C_{n+1}^{(n+1)/2} \quad if \quad \text{$(n + 1)\, \vdots\, 4$}; \\
8\Large \sum\limits_{k=0}^{(n - 1)/4}\dfrac{C_{n+1}^{2k}}{n - 4k + 1} \Large \sum\limits_{l=0}^{(n - 4k - 1)/2} \dfrac{1}{2k + 1 + 2l} - \\ - 8\Large \sum\limits_{k=0}^{(n - 5)/4}\dfrac{C_{n+1}^{2k+1}}{n - 4k - 1} \Large \sum\limits_{l=0}^{(n - 4k - 3)/2} \dfrac{1}{2k + 3 + 2l} + \\ + 4 C_{n+1}^{(n+1)/2} \Large \sum\limits_{l=0}^{(n - 1)/4} \dfrac{1}{(2l+1)^2} - \\ - \dfrac{\pi^2}{2} C_{n+1}^{(n+1)/2} \quad if \quad \text{$(n - 1)\, \vdots\, 4$}. 
\end{cases}
\]
Here the sign $\vdots$ denotes the divisibility of the first number by the second.


\begin{thebibliography}{00}
%
\bibitem{lib1} G.~Aad et al. Observation of a new particle in the search for the Standard Model Higgs boson with the ATLAS detector at the LHC.  \textit{Phys. Lett. B}  {\bf 716}, 1 (2012), 
doi:10.1016/j.physletb.2012.08.020

\bibitem{lib2} S.~Chatrchyan et al. Observation of a New Boson at a mass of 125 GeV with the CMS experiment at the LHC. \textit{Phys. Lett. B} {\bf 716}, 30 (2012), doi:10.1016/j.physletb.2012.08.021

\bibitem{lib3} D.~de Florian et al. \textit{Handbook of LHC Higgs Cross Sections: 4. Deciphering the Nature of the Higgs Sector, volume 2 (2017)} (CERN, 2017) (ISBN 978-92-9083-443-4).

\bibitem{lib4} N.~Cabibbo. Unitary symmetry and leptonic decays. \textit{Phys. Rev. Lett.} {\bf 10}, 531 (1963), doi:10.1103/PhysRevLett.10.531

\bibitem{lib5} Makoto~Kobayashi, Toshihide~Maskawa. CP violation in the renormalizable theory of weak interaction. \textit{Prog. Theor. Phys.} {\bf 49}, 652 (1973), doi:10.1143/PTP.49.652

\bibitem{lib6} G.~R.~Farrar, M.~E.~Shaposhnikov.  Baryon asymmetry of the universe in the standard 
electroweak theory.  \textit{Phys. Rev. D} {\bf 50}, 774 (1994), doi:10.1103/PhysRevD.50.774

\bibitem{lib7} S.~Davidson, E.~Nardi, Y.~Nir. Leptogenesis.  \textit{Phys. Rept.} {\bf 466}, 105 (2008), 
doi:10.1016/j.phys-rep.2008.06.002

\bibitem{Bass:2021} S.D.~Bass, A.~De Roeck and M.~Kado. The Higgs boson implications and prospects for future discoveries. \textit{Nat. Rev. Phys.} {\bf 3}, 608 -- 624 (2021), doi:10.1038/s42254-021-00341-2

\bibitem{CMS:2012vby} S.~Chatrchyan et al. [CMS]. Study of the Mass and Spin-Parity of the Higgs Boson Candidate Via Its Decays to Z Boson Pairs. \textit{Phys. Rev. Lett.} {\bf 110}, no.8, 081803 (2013),
doi:10.1103/PhysRevLett.110.081803

\bibitem{CMS:2014nkk}
V.~Khachatryan et al. [CMS]. Constraints on the spin-parity and anomalous HVV couplings of the Higgs boson in proton collisions at 7 and 8 TeV. \textit{Phys. Rev. D} {\bf 92}, no.1, 012004  (2015),
doi:10.1103/PhysRevD.92.012004

\bibitem{ATLAS:2013xga} G.~Aad et al. [ATLAS]. Evidence for the spin-0 nature of the Higgs boson using ATLAS data. \textit{Phys. Lett. B} {\bf 726}, 120-144 (2013),
doi:10.1016/j.physletb.2013.08.026

\bibitem{CMS:2019rvj}
A.~M.~Sirunyan et al. [CMS]. Search for production of four top quarks in final states with same-sign or multiple leptons in proton-proton collisions at $\sqrt{s}=$ 13 TeV. 
\textit{Eur. Phys. J. C} {\bf 80}, no.2, 75 (2020),
doi:10.1140/epjc/s10052-019-7593-7

\bibitem{ATLAS:2020ior} G.~Aad et al. [ATLAS]. $CP$ Properties of Higgs Boson Interactions with Top Quarks in the $t\bar{t}H$ and $tH$ Processes Using $H \rightarrow \gamma\gamma$ with the ATLAS Detector. \textit{Phys. Rev. Lett.} {\bf 125}, no.6, 061802 (2020), \ doi:10.1103/PhysRevLett.125.061802

\bibitem{CMS:2021nnc}
A.~M.~Sirunyan et al. [CMS]. Constraints on anomalous Higgs boson couplings to vector bosons and fermions in its production and decay using the four-lepton final state. \textit{Phys. Rev. D} {\bf 104}, no.5, 052004 (2021),
\ doi:10.1103/PhysRevD.104.052004

\bibitem{Berge:2011ij} S.~Berge, W.~Bernreuther, B.~Niepelt and H.~Spiesberger. How to pin down the CP quantum numbers of a Higgs boson in its tau decays at the LHC. \textit{Phys. Rev. D} {\bf84}, 116003 (2011),
\ doi:10.1103/PhysRevD.84.116003

\bibitem{Bower:2002zx}
G.R.~Bower, T.~Pierzchala, Z.~Was, M.~Worek.
Measuring the Higgs boson's parity using tau $\to $ rho nu.    \textit{Phys. Lett. B} {\bf 543}, 
227-234 (2002), \  doi:10.1016/S0370-2693(02)02445-0    

\bibitem{Desch:2003mw}
K.~Desch, Z.~Was, M.~Worek. Measuring the Higgs boson parity at a linear collider using the tau 
impact parameter and tau $\to$ rho nu decay.  \textit{Eur. Phys. J. C} {\bf 29}, 491-496 (2003), \
doi:10.1140/epjc/s2003-01231-4

\bibitem{Desch:2003rw}
K.~Desch, A.~Imhof, Z.~Was, M.~Worek.
Probing the CP nature of the Higgs boson at linear colliders with 
tau spin correlations: The Case of mixed scalar-pseudoscalar couplings.
\textit{Phys. Lett. B} {\bf 579}, 157-164 (2004), \ doi:10.1016/j.physletb.2003.10.074

\bibitem{Barberio:2017ngd}
E.~Barberio, B.~Le, E.~Richter-Was, Z.~Was, D.~Zanzi, J.~Zaremba. 
Deep learning approach to the Higgs boson CP measurement in 
$H \to \tau \tau$ decay and associated systematics.
\textit{Phys. Rev. D} {\bf 96}, 7, 073002 (2017), \  doi:10.1103/PhysRevD.96.073002
   
\bibitem{Lasocha:2020ctd}
K.~Lasocha, E.~Richter-Was, M.~Sadowski, Z.~Was.
Deep neural network application: Higgs boson $CP$ state mixing 
angle in $H \to \tau\tau$ decay and at the LHC.
\textit{Phys. Rev. D} {\bf 103}, 3, 036003 (2021),  \ doi:10.1103/PhysRevD.103.036003

\bibitem{Bernreuther:1997} W.~Bernreuther, A.~Brandenburg, and M.~ Flesch. QCD corrections to decay distributions of neutral Higgs bosons with (in)definite CP parity. \textit{Phys. Rev. D} {\bf 56}, 90 (1997), \
doi:10.1103/PhysRevD.56.90

\bibitem{lib8} A.~Yu.~Korchin, V.~A.~Kovalchuk. Decay of the Higgs boson to $\tau^-$ $\tau^+$ and non-Hermiticity of the Yukawa interaction. \textit{Phys. Rev. D} {\bf 94}, 076003 (2016), \
doi:10.1103/PhysRevD.94.076003

\bibitem{PDG} R.~L.~Workman et al. (Particle Data Group), Review of Particle Physics. 
\textit{PTEP} {\bf 2022}, 083C01 (2022),  \ doi:10.1093/ptep/ptac097 

\bibitem{Braaten:1980}
E. Braaten, J.~P. Leveille.  Higgs Boson Decay and the Running Mass. 
\textit{Phys. Rev. D}  {\bf 22}, 715 (1980), \ doi:10.1103/PhysRevD.22.715

\bibitem{Janot:1989}
Patrick Janot. First order QED and QCD radiative corrections to Higgs decay into massive fermions.
\textit{Phys. Lett. B} {\bf 223}, 110-118 (1989),  \ doi:10.1016/0370-2693(89)90929-5

\bibitem{Drees:1990}
Manuel Drees, Ken-ichi Hikasa. Note on QCD corrections to hadronic Higgs decay. 
\textit{Phys. Lett. B} {\bf 240}, 455 (1990),  \textit{Phys. Lett. B} {\bf 262}, 497 (1991) (erratum), \
doi:10.1016/0370-2693(90)91130-4

\bibitem{Kalyniak:1991}
Pat Kalyniak, Nita Sinha, Rahul Sinha, John N. Ng. QCD corrections to Higgs boson 
decay and jet analysis. \textit{Phys. Rev. D} {\bf 43}, 3664-3668 (1991), \ doi:10.1103/PhysRevD.43.3664

\bibitem{Gorishnii:1990zu}
S.~G.~Gorishnii, A.~L.~Kataev, S.~A.~Larin and L.~R.~Surguladze.
Corrected Three Loop QCD Correction to the Correlator of the
Quark Scalar Currents and $\Gamma_{tot}$($H^0 \to$ hadrons).
\textit{Mod. Phys. Lett. A} \textbf{5}, 2703-2712 (1990), \ doi:10.1142/S0217732390003152

\bibitem{Gorishnii:1991zr}
S.~G.~Gorishnii, A.~L.~Kataev, S.~A.~Larin and L.~R.~Surguladze.
Scheme dependence of the next to next-to-leading QCD corrections to
$\Gamma_{tot}$($H^0 \to $ hadrons) and the spurious QCD
infrared fixed point. \textit{Phys. Rev. D} \textbf{43}, 1633-1640 (1991), \ doi:10.1103/PhysRevD.43.1633

\bibitem{Kataev:1993be}
A.L.~Kataev and V.T.~Kim. 
The Effects of the QCD corrections to $\Gamma$($H^0 \to b \bar{b}$).
\textit{Mod. Phys. Lett. A} \textbf{9}, 1309-1326 (1994),  \ doi:10.1142/S0217732394001131

\bibitem{Kataev:2008ntk}
A.~L.~Kataev and V.~T.~Kim.
Uncertainties of QCD predictions for Higgs boson decay into bottom quarks at NNLO and beyond.
\textit{PoS} \textbf{ACAT08}, 004 (2008),  \ doi:10.22323/1.070.0004

\bibitem{lib9} V.~B.~Berestetskii, L.~P.~Pitaevskii, E.~M.~Lifshitz. \textit{Quantum Electrodynamics. Vol. 4, 2nd Edition}  (Butterworth-Heinemann, 1982) [ISBN: 9780080503462].

\bibitem{lib10} F.~J.~Yndurain. \textit{Quantum Chromodynamics.	An Introduction to the Theory of Quarks and Gluons.} (Springer Berlin, Heidelberg, 1983) [ISBN: 978-3-662-09633-8]. 

\bibitem{lib11} A.~Yu.~Korchin, V.~A.~Kovalchuk. Decay of the Higgs boson $ h \rightarrow \tau^- \tau^+ \rightarrow \pi^- \nu_\tau \pi^+ \bar \nu_\tau$ for a non-Hermitian Yukawa interaction. 
\textit{Acta Phys. Polon. B} {\bf 53}, 1 (2022), \ doi:10.5506/APhysPolB.53.1-A2 

\bibitem{Kawasaki:1973} S.~Kawasaki, T.~Shirafuji, S.~Y.~Tsai.
Productions and decays of short-lived particles in $e^{+} e^{-}$ colliding beam experiments.
\textit{Prog. Theor. Phys.} {\bf 49},  1656--1678  (1973), \
doi:10.1143/PTP.49.1656  

\bibitem{Grzadkowski:1995} B.~Grzadkowski, J.~F.~Gunion. Using decay angle correlations to detect 
CP violation in the neutral Higgs sector. \textit{Phys. Lett. B} {\bf 350}, 218-224 (1995),  \
doi:10.1016/0370-2693(95)00369-V

\end{thebibliography}
\end{document}